\documentclass[a4paper,fleqn,usenatbib]{mnras}

\usepackage{newtxtext,newtxmath,deluxetable}

\usepackage[T1]{fontenc}
\usepackage{ae,aecompl}

\usepackage{graphicx}	
\usepackage{amsmath}	
\usepackage{amssymb}	
\usepackage[english]{babel}
\usepackage{array}
\usepackage{multirow}

\newcolumntype{M}[1]{>{\centering\arraybackslash}m{#1}}

\title[Occultation in NGC 3227]{A rapid occultation event in NGC 3227}

\author[T.J.Turner et al.]{
T.J. Turner,$^{1,2}$\thanks{E-mail: tjturner@umbc.edu}
J.N. Reeves,$^{2}$
V. Braito,$^{2,4}$
A Lobban$^3$
S. Kraemer$^{5}$ 
and L. Miller$^{6}$ 
\\
$^{1}$Department of Physics, University of Maryland Baltimore County, 
   Baltimore, MD 21250 U.S.A\\
$^{2}$Center for Space Science and Technology, University of Maryland Baltimore County, 1000 Hilltop Circle, Baltimore, MD 21250, USA\\
$^{3}$Astrophysics Group, School of Physical and Geographical Sciences, Keele 
University, Keele, Staffordshire ST5 5BG, U.K\\
$^{4}$INAF - Osservatorio Astronomico di Brera, Via Bianchi 46 I-23807 Merate (LC), Italy\\
$^5$ The Catholic University of America, Washington, DC 20064, USA \\
$^6$ Department of Physics, University of Oxford, Keble Road, Oxford, OX1 3RH, U.K. \\
 }

\date{Accepted XXX. Received YYY; in original form ZZZ}

\pubyear{2017}

\begin{document}
\label{firstpage}
\pagerange{\pageref{firstpage}--\pageref{lastpage}}
\maketitle

\begin{abstract}

NGC~3227 exhibits rapid  flux and spectral variability in the X-ray band. To understand this behaviour we conducted a co-ordinated observing campaign using 320 ks of {\it XMM-Newton} exposures together with 160 ks of overlapping  {\it NuSTAR} observations, spanning a month. Here, we present a rapid variability event that occurs toward the end of the campaign. The spectral hardening event  is accompanied by  a change in the depth of an unresolved transition array,  whose time-dependent behaviour is resolved using the RGS data. 
This UTA fingerprint allows us to identify this as a transit event, where a clump of gas having $N_H \sim 5 \times 10^{22}\,{\rm atoms\, cm^{-2}}$, log $\xi \sim 2$ occults $\sim 60\%$ of the continuum photons over the course of approximately a day.  This occulting gas is likely associated with clouds in the inner BLR. An additional zone of gas with lower column and higher ionization, matches the outflow velocity of the variable zone, and may represent transmission through the cloud limb. 

\end{abstract}

\begin{keywords}
galaxies:active -- galaxies: individual: NGC 3227 -- galaxies: Seyfert -- X-rays: galaxies
\end{keywords}



\section{Introduction}

Detailed measurement of X-ray spectral features, especially using grating data, has revealed a complex, multi-zoned outflowing X-ray reprocessor that shapes the observed properties of Active Galactic Nuclei \citep[e.g.][]{blustin05a}.  X-ray  observations are particularly valuable for understanding the mass and energy carried by the gas, as they can trace the entire radial extent of the nuclear outflow. 
Fluctuations in the intrinsic  X-ray continuum will reverberate from the reprocessing gas, which may show a delayed scattered X-ray signal or a change in gas ionization state in response to the continuum. The detectability of these responses depends on many factors, including the degree of continuum variation, initial state of the gas and the orientation of the system. For cases where the plane of the reprocessing gas lies predominantly in our line-of-sight, 
one might expect to see occultation and uncovering events  if the reprocessor is clumpy, such as in the form of a cloud ensemble. 

The X-ray reprocessing zones have outflow velocities that range from hundreds of km/s to a fraction of c (Tombesi et al. 2013, Gofford et al. 2013) and may be phenomenologically linked to the outflows observed in the UV and optical spectra of these sources. The gas feeding the flow may originate from the surface of the accretion disk, the optical broad line clouds, the putative obscuring torus or may include contributions from all of these.

\begin{table*}
\centering
\caption{Observation Log}
\begin{tabular}{| l | c | c | c | }
\hline

Date & 
Observation ID  & 
Total & 
Flux \\
 
 & 
 & 
Exposure & 
2-10\,keV  \\
\hline
\hline
{\it XMM}  &   & \\
\hline
2016-11-09 12:51:03 & 0782520201  & 92.0  & 3.24 \\
2016-11-25  10:25:35 & 0782520301   & 74.0   & 2.59   \\
2016-11-29  13:10:38 & 0782520401  & 84.0  &   3.12  \\
2016-12-01  09:58:49 & 0782520501  & 87.0  &  3.65  \\
{\bf 2016-12-05  09:42:23} & {\bf 0782520601} & {\bf 86.6}  &  {\bf 4.10}  \\
{\bf 2016-12-09  09:24:49} & {\bf 0782520701}  & {\bf 87.9}  &  {\bf 3.78}  \\
\hline
{\it NuSTAR} &   &  & 10 - 50 keV \\
\hline
2016-11-09 13:16:08  & 60202002002 &   49.8  & 6.99  \\
2016-11-25   09:26:08 & 60202002004 &  42.5  & 5.99  \\
2016-11-29  16:31:08 & 60202002006 & 39.7  & 6.37 \\
2016-12-01  10:31:08 & 60202002008 &  41.8  & 7.47 \\
{\bf 2016-12-05   09:31:08} &  {\bf 60202002010} & {\bf 40.9}  & {\bf 7.79}  \\
{\bf 2016-12-09  08:36:08} & {\bf 60202002012}   &  {\bf 39.2}  & {\bf 7.56}  \\
\hline
\end{tabular} \\
\vspace{-0.7cm}
\tablecomments{Observed fluxes are given in units 10$^{-11}$ erg\,cm$^{-2}{\rm s^{-1}}$ based on the {\it XMM} pn  and the mean FPM data. Bold face type denotes the sequences presented in this paper.}
\label{tab:data}
\end{table*}%

 For such complex systems,  simultaneous consideration of spectral and timing 
properties is necessary to disentangle the various modes of behaviour and elucidate the location, physical state, mass and energy carried in the reprocessing gas.  
Long timescale absorption events have been observed in sources such as NGC~3783, whose variability reveals a clumpy X-ray reprocessing wind   \citep[e.g.][]{mehdipour17a} located at the outer part of the optical broad-line region  (BLR).  The X-ray light curves of MCG-6-30-15 \citep{mckernan98a} and NGC 3516 \citep{turner08a} have been captured during 
 dips whose profiles are characteristic of occultation events, also indicative of absorption close to BLR radii.  Similar X-ray absorption events have also been found in Mrk~335 \citep{longinotti13a}, NGC 5548 \citep{kaastra14a} and NGC 985 \citep{ebrero16a}. 

X-ray spectral data have been particularly valuable, as they can trace variability in spectral features, such as in  the Fe K-band absorption lines as seen in  NGC 1365 \citep{risaliti07a, risaliti09b} and NGC 3227 \citep{lamer03a} - unambiguous signatures of  changes in the line-of-sight opacity on  timescales of days.   In the most extensively studied sources, such as NGC 1365 \citep{braito14a}, NGC 5548  \citep{cappi16a} and NGC 3783 \citep{kaastra18a} absorption variations are observed across a wide range of timescales,  showing a large radial range for the X-ray absorption complex,  down to the accretion disc scale \citep{kaastra14a}. It seems likely that the X-ray reprocessor is composed of zones covering a wide range of radii such that variability result captured depends on the observation specifics convolved with the scale size and physical state of the gas in a particular AGN.

More generally, the characteristic softening of the observed X-ray spectrum in high flux states is very common in AGN. This behaviour has been successfully modeled as  absorption changes, such as variations in the gas covering fraction such that high states contain a greater fraction of direct light than low states (e.g. MCG-06-30-16, \citealt{miller08a};  Mrk~766, \citealt{miller07a}; NGC~3227, \citealt{lamer03a}).  In a number of targets, X-ray variability timescales and fitted ionization parameters have been among measurements used to determine that a significant component of the X-ray reprocessing gas  is likely co-incident with the BLR \citep[e.g.][]{bianchi12a}. 

Thus, results to date are building up a new Unified Model for AGN, built around the idea of a cloud ensemble as the obscuring material. A concentration of clouds is suggested to lie close to the equatorial plane of the accretion disk, with an angular distribution that falls off toward the poles 
\citep{nenkova02a, nenkova08a, elitzur06a}. New infrared data yield information on the gas distribution outside of the dust sublimation radius and results generally support a cloud ensemble rather than a simple torus distribution for the gas 
\citep[e.g.][]{alonso-Herrero11a}. 

NGC 3227 is a Seyfert 1.5 galaxy at z=0.003859,  that has shown high unabsorbed \citep[e.g.][]{lamer03a} and absorbed low X-ray states in previous X-ray observations \citep{rivers11a,markowitz09a,markowitz14a}. 
 A  study using  {\it  XMM}, {\it Suzaku} and {\it Swift} data  \citep{beuchert15a} found an outflowing warm absorber that is complex, with signatures from three zones of ionized gas with transit by one inhomogeneous clump of gas linking part of the absorber complex to the location of the optical broad-line region. Further to this, variations have been observed in the Fe K-shell absorption lines (Gofford et al 2013) and a so-called negative lag detected in the lag-frequency spectrum, indicating the presence of a significant reveberated signal from reprocessing gas out of the line-of-sight \citep{demarco13a}.  
 
Time-resolved absorption events, such as that reported by  \citep{beuchert15a}, are still quite rare. The observation of such in NGC 3227, along with its wide range of observed spectral states motivated the  
2016 {\it XMM}/{\it NuSTAR} campaign comprising $\sim$320 ks of  {\it XMM-Newton} time overlapping  $\sim 160$ ks of {\it NuSTAR} observations. The new campaign sampled the source on days to weeks to elucidate the nature of the X-ray reprocessor.  In this paper we present results from a sub-set of the new data, within which a rapid absorption event was isolated.  

In Section 2 we provide details of the observations and data reduction.  In Section 3 
we present a brief overview of the campaign, showing the source light curve and the hardness ratio behaviour that identifies the time period of particular interest with regard to spectral variability. In Section 4 we develop a spectral model for the source prior to the absorption event. In Section 5 we assess the spectral variability of NGC 3227 from a subset of observations from 2016 Dec 05-09, using both the RGS and 
pn spectra to quantify the absorber variability. In Section 6 we discuss the parameters drawn from our results and attempt to place the relative location of the absorbing gas. 

\section{Observations}

\subsection{XMM-Newton}

{\it XMM-Newton}  (hereafter {\it XMM}) conducted six observations of  NGC 3227 over the period 2016 Nov 09 - Dec 09 as part of a co-ordinated 
{\it XMM}/{\it NuSTAR} campaign).  The  {\it XMM} observations were spaced over timescales of days to weeks to sample the source's different flux states, with exposures between 74 and 92 ks (Lobban et al 2018, in prep).   Table~1 gives a summary of the observations. The 2016 Dec 09 data from {\sc OBSID} 0782520701 captured a rapid absorption event (Figures~\ref{fig:pn_lc}, \ref{fig:hardness}). Here we present the {\it XMM} spectral data that allow a parameterization of the event. To set the baseline for the 
spectral change, we also include spectral analysis of the preceding data from 2016 Dec 05 ({\sc OBSID} 0782520601,  
along with the overlapping {\it NuSTAR} data.

All the {\it XMM} observations  were performed in Small Window mode, with the medium filter applied. All data were processed using  {\sc sas} v16.1.0 and {\sc heasoft} v6.23 software and the data reduction and extraction is described in full in a companion paper by Lobban et al (2018, in prep). 
Each observation was  filtered  for high background, which  did not affect the subset of observations presented here.   We extracted the  EPIC pn  source and background  spectra  using a
circular region  with a   radius of $35''$  and    two circular regions  with a radius of $28''$, respectively.  After cleaning, the Dec 05 (601) and Dec 09 (701) {\it XMM} events files yielded exposures of $\sim 60$ ks and 53 ks respectively, with  mean 0.5-10.0 keV count rates of $14.5$ and 10.3 source ct/s, in the pn and 0.36 and 0.23 ct/s, respectively over 0.5 -2.0 keV in the summed RGS 1 and 2 spectra for the two dates.  We generated the response matrices and  the ancillary response files at the source position   using the SAS tasks \textit{arfgen} and \textit{rmfgen} and the latest calibration available. The pn spectrum was  binned to a minimum of 50 counts per energy bin: the binned data maintained sampling finer than the spectral resolution of the pn. The background rate was  $< 1$\% of the net source rate in the pn. 

The {\it XMM} RGS1 and RGS2 data were reduced  with  the standard {\sc sas} task {\sc rgsproc}, where we filtered   for  high background time intervals  applying a threshold of 0.2\,cts\,s$^{-1}$ on the background event files.  We then combined the RGS1 and RGS2 spectra with the {\sc sas} task {\sc rgscombine}, after checking that  the  RGS1 and RGS2 spectra were in good agreement. 
We then extracted the background-corrected light curves   using the {\sc sas} task {\it rgslccorr} for the total  RGS band  adopting a binsize of 1 ks. 
 We inspected  the light curves obtained for each RGS (considering only the first order data) as well as for the whole RGS (i.e. combining both the RGS1 and RGS2) and we  found  that  during the Dec 09 observation  NGC3227 clearly varied also in the RGS band. We then extracted and combined the  RGS1 and RGS2   spectra    for each of the time intervals that we defined (see Section 3).  
 The RGS spectra  collected during the sub-slices of 0782520701 (see Section 3)  received a constant  spectral binning,  $\Delta \lambda=0.1 $\AA\, 
while  for the  averaged RGS1 and 2 spectra from each of  sequences 0782520601, 0782520701 received a 
finer binning,  ($\Delta \lambda=0.03$\AA). The background comprised $< 7\%$ of the net source rate in the RGS data over 0.5 - 2 keV; note   that  even at the finer binning each of the RGS the spectral bins have more than 20 counts per bin,  which allow us to use the $\chi^2$ statistic  for the spectral fitting.

\subsection{NuSTAR}

{\it NuSTAR}  carries  two co-aligned telescopes containing Focal Plane Modules A and B (FPMA, FPMB; \citealt{harrison13a} ) covering a useful bandpass of $\sim 3-80$ keV for AGN. 
{\it NuSTAR} observed NGC 3227 during 2016 Nov 09 - Dec 09 covering the  {\it XMM} baseline with overlapping exposures occurring on the same days, as required for the joint observational campaign with {\it XMM}.  An additional observation was made 2017 Jan 21 to co-ordinate with a GTO observation performed using {\it Chandra} HETG. Those data are not included here as they occurred approximately six weeks after the end of the nominal campaign, but they will be included in a future paper. 
The {\it NuSTAR} {\sc OBSID} identifiers for the exposures from 2016 Dec 05 and Dec 09  presented here are 60202002010 and 60202002012. 

Event files were created through the {\sc nupipeline} task, calibrated with files from {\sc caldb  20180419} 
and cleaned  applying  the standard screening criteria, where we filtered for the passages through the SAA  setting the mode    to  ``optimised"  in \textsc{nucalsaa} . This  yields  net exposures  of $\sim 40$ ks per focal plane module.   For each of the Focal Plane Module (FPMA and FPMB)  the source spectra were extracted  from a circular region with a radius of $70''$,  while the background spectra were extracted  from a circular region with a $75''$ radius located on the same detector.  
NuSTAR source spectra were  binned to  100 counts per spectral channel, maintaining a sampling that is 
finer than the spectral resolution of the instruments.  (In some of the plots the data are binned more coarsely than allowed in the fit, for visual clarity.) 

During 2016 Dec 05 FPMA  gave a mean source count rate over 3-50 keV of  $1.18\pm 0.005$ ct/s, FPMB gave  $1.12\pm0.005$ ct/s.  Dec 09 FPMA yielded $1.14\pm0.007$, FPMB gave $1.06\pm0.005$; the background level was $\sim 1\%$ of the total count rate for both dates and modules.
These rates correspond to fluxes  $\sim 4.0 \times 10^{-11} {\rm erg\, cm^{-2} s^{-1}}$ in the 2-10 keV band and 
$\sim 7.5 \times 10^{-11} {\rm erg\, cm^{-2} s^{-1}}$  in the 10 - 50 keV band, for both epochs.

There is a small  calibration offset recommended to be allowed between FPMA and FPMB 
and so a constant component was allowed in all models, constrained to a range of 0.9-1.1 for the cross-normalization constant between the pn and {\it NuSTAR} detectors. 

\subsection{General Considerations}

Spectra are analyzed with XSPEC v12.9.1m.  We used data over 0.5 - 10 keV for the pn, 0.4-2.0 keV for RGS  and 3 - 70 keV for {\it NuSTAR}. 
All models included the Galactic line-of-sight absorption,
N$_{\rm H,Gal} = 2.0 \times 10^{20}{\rm cm}^{-2}$ \citep{dickey90a}, parameterized using {\sc tbabs}.  All model
components were adjusted to be at the redshift of the host galaxy,
except for the Galactic absorption. 
 For the
ionized absorber model tables we used version 2.41 of the {\sc xstar} code
\citep{kallman01a,kallman04a}, assuming the abundances of
\citet{grevesse98a}. {\sc xstar} models the absorbing gas as thin
slabs, with parameters of atomic column density and ionization parameter $\xi$,
defined as $$\xi=\frac{L_{\rm ion}}{n_e r^2}$$ that has units erg\,cm\,s$^{-1}$
and where $L_{\rm ion}$ is the ionizing luminosity between 1 and
1000 Rydbergs, $n_e$ is the gas density in cm$^{-3}$ and $r$ is the
radial distance (cm) of the absorbing gas from the central continuum
source. The spectral energy distribution was taken to be a simple
power law with $\Gamma=2$. The turbulent
velocity was taken as $\sigma=300$ km s$^{-1}$. (The turbulent velocity widths are consistent with the 
upper limits on the narrow absorption lines seen later in the RGS data.)

In the fits, unless otherwise
stated, parameters are quoted in the rest-frame of the source and
errors are at the 90\% confidence level for one interesting parameter
($\Delta \chi^2 = 2.706$).

\begin{figure*}
\includegraphics[scale=0.43,width=16cm, height=8cm,angle=0]{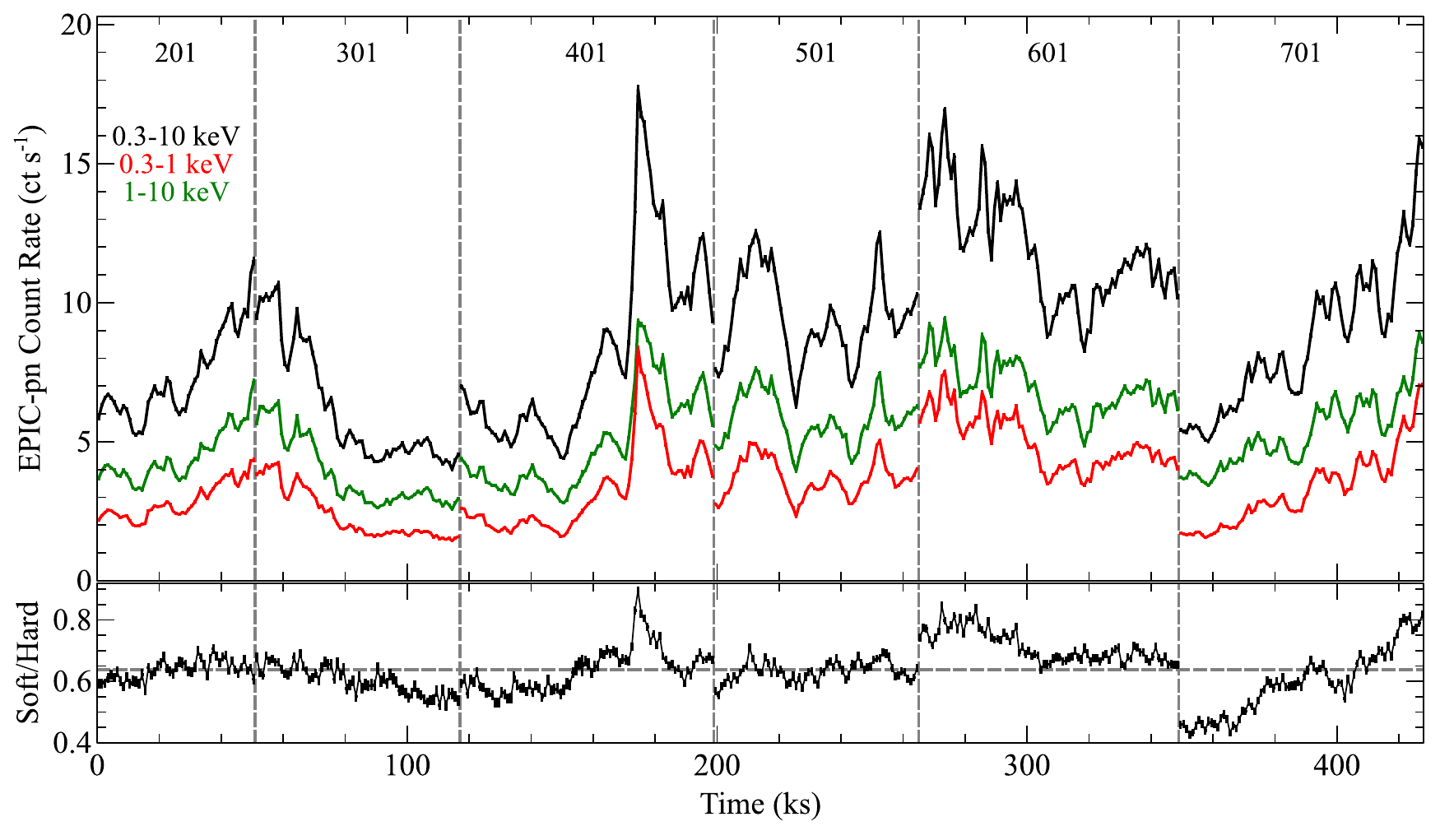}
\caption{Upper panel: the {\it XMM-Newton} EPIC-pn lightcurve of NGC\,3227 in 1\,ks bins across all six observations from 2016.  The broad-band lightcurve (0.3-10\,keV) is shown in black while soft (0.3-1\,keV) and hard (1-10\,keV) bands are shown in red and green, respectively.  Lower panel: the evolution of the `softness ratio' (i.e. soft/hard) over the course of the campaign.
} 
\label{fig:pn_lc}
\end{figure*}

\begin{figure}
\includegraphics[scale=0.43,width=8cm, height=8cm,angle=0]{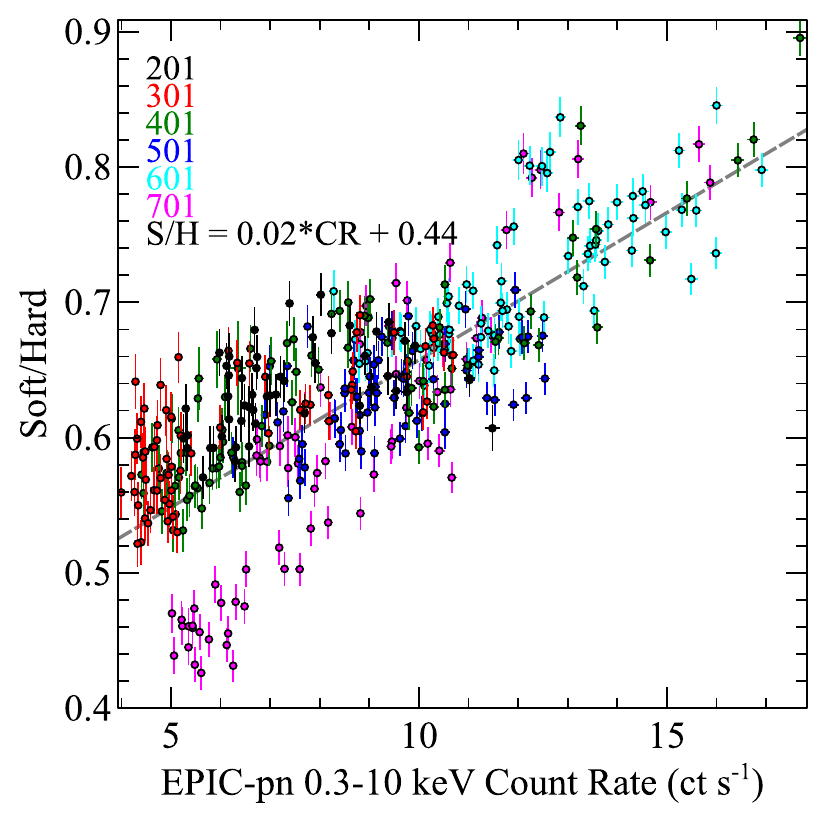}
\caption{The `softness ratio' (i.e. soft/hard) of NGC\,3227 using the EPIC-pn data in 1\,ks bins.  The six observations are shown in different colours - chronologically: black, red, green, blue, cyan, magenta.  The soft and hard bands are defined as 0.3-1 and 1-10\,keV, respectively.  The grey dashed line shows a linear best-fitting model with a slope of 0.02 and a positive offset of 0.44.  Note that observation 701 (magenta) lies off the best-fitting slope, hinting at unusual spectral behaviour during this time.
} 
\label{fig:hardness}
\end{figure}

\begin{figure}
\includegraphics[scale=0.43,width=8cm, height=8cm,angle=-90]{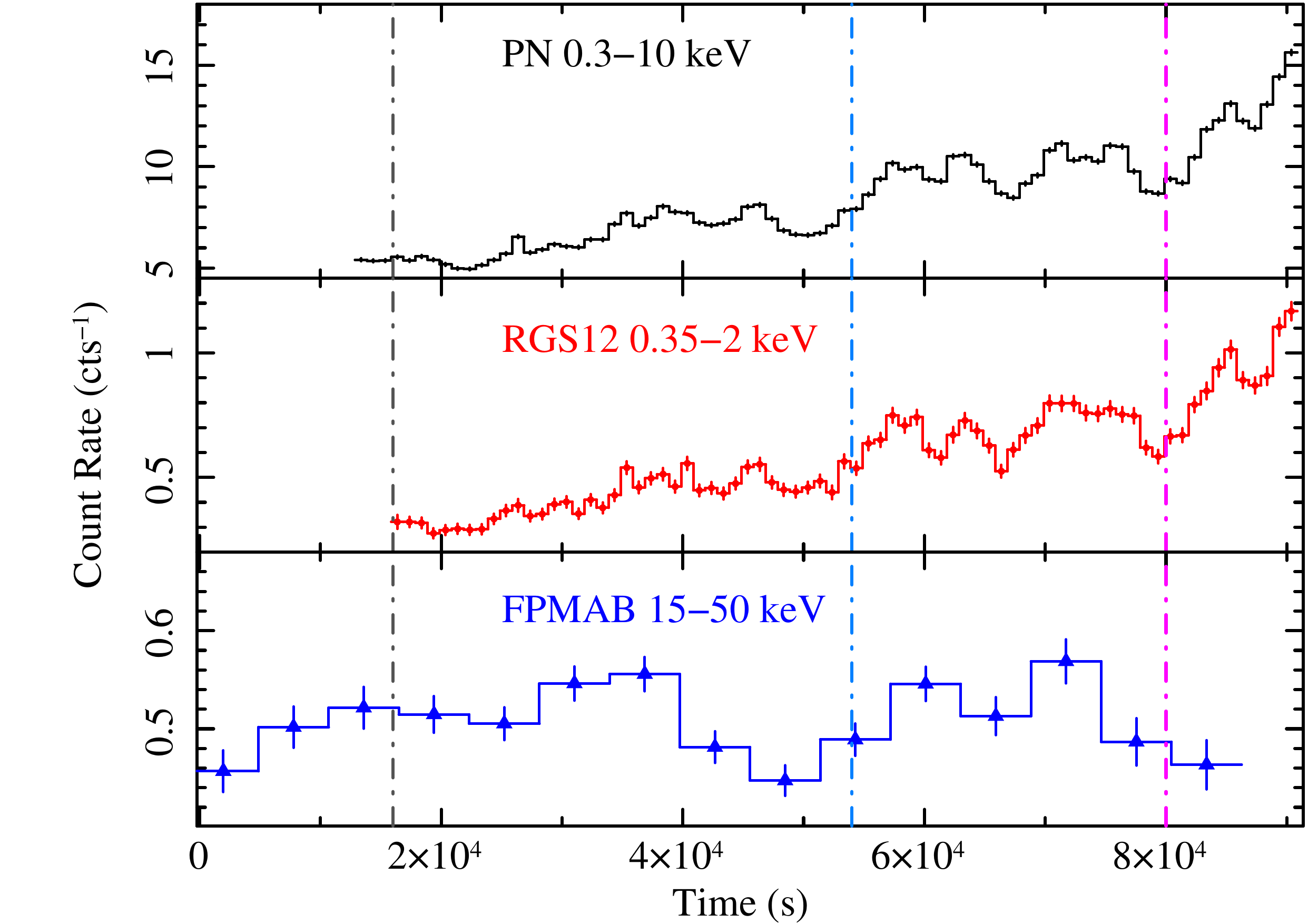}
\caption{The light curves for {\it XMM} RGS and pn instruments, using 1000 s time bins  and the overlapping {\it NuSTAR} data from Dec 09, binned to the {\it NuSTAR} orbit timescale ($\sim 5800 $s).  The vertical dash-dot lines show the time selections: the first interval is discarded as it contains only {\it NuSTAR} data without {\it XMM}, the second interval is denoted slice A, third is slice B and the final slice is C.}
\label{fig:pn_rg}
\end{figure}

\begin{figure}
\includegraphics[scale=0.43,width=6cm, height=8cm,angle=-90]{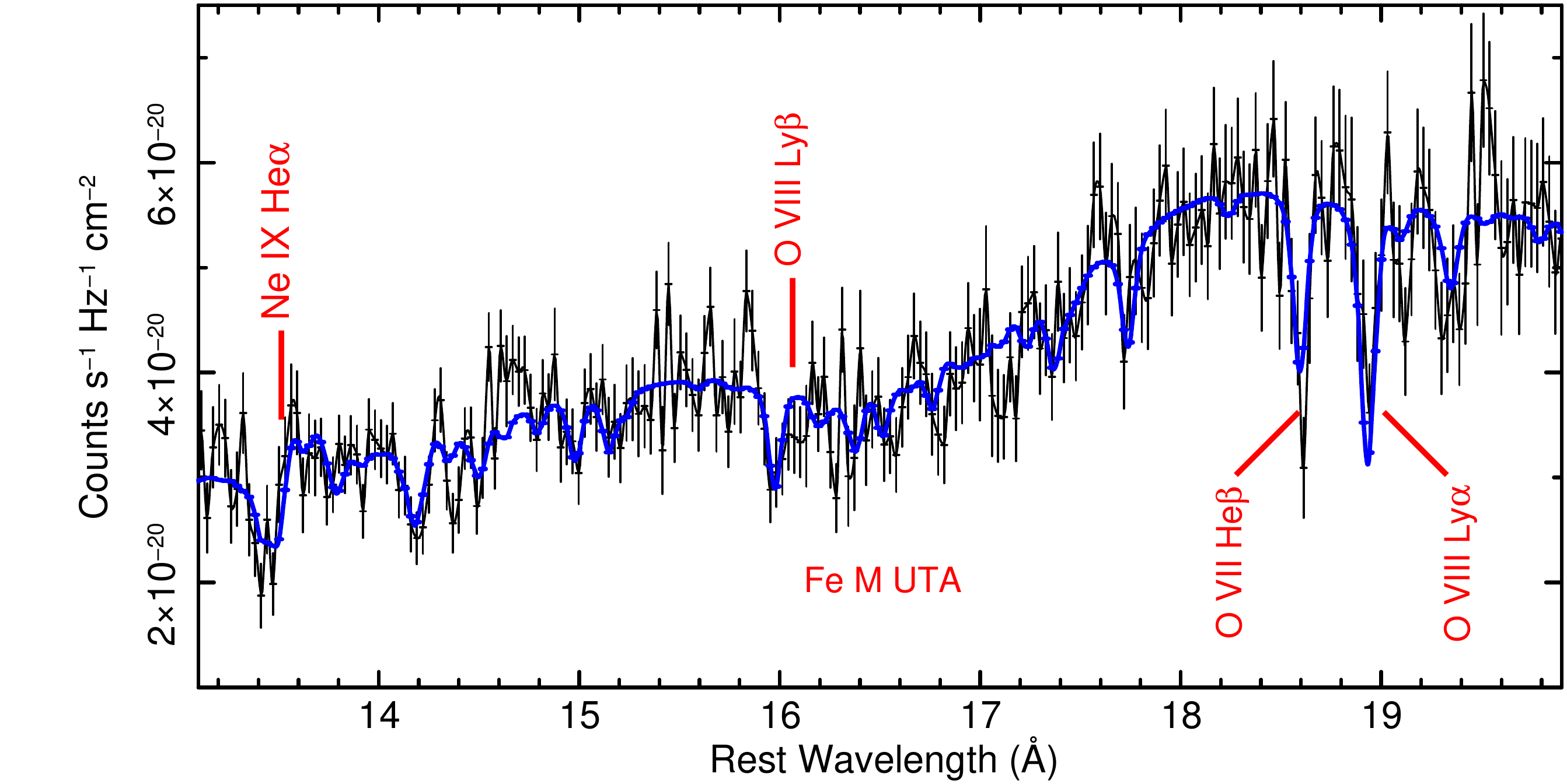}
\includegraphics[scale=0.43,width=6cm, height=8cm,angle=-90]{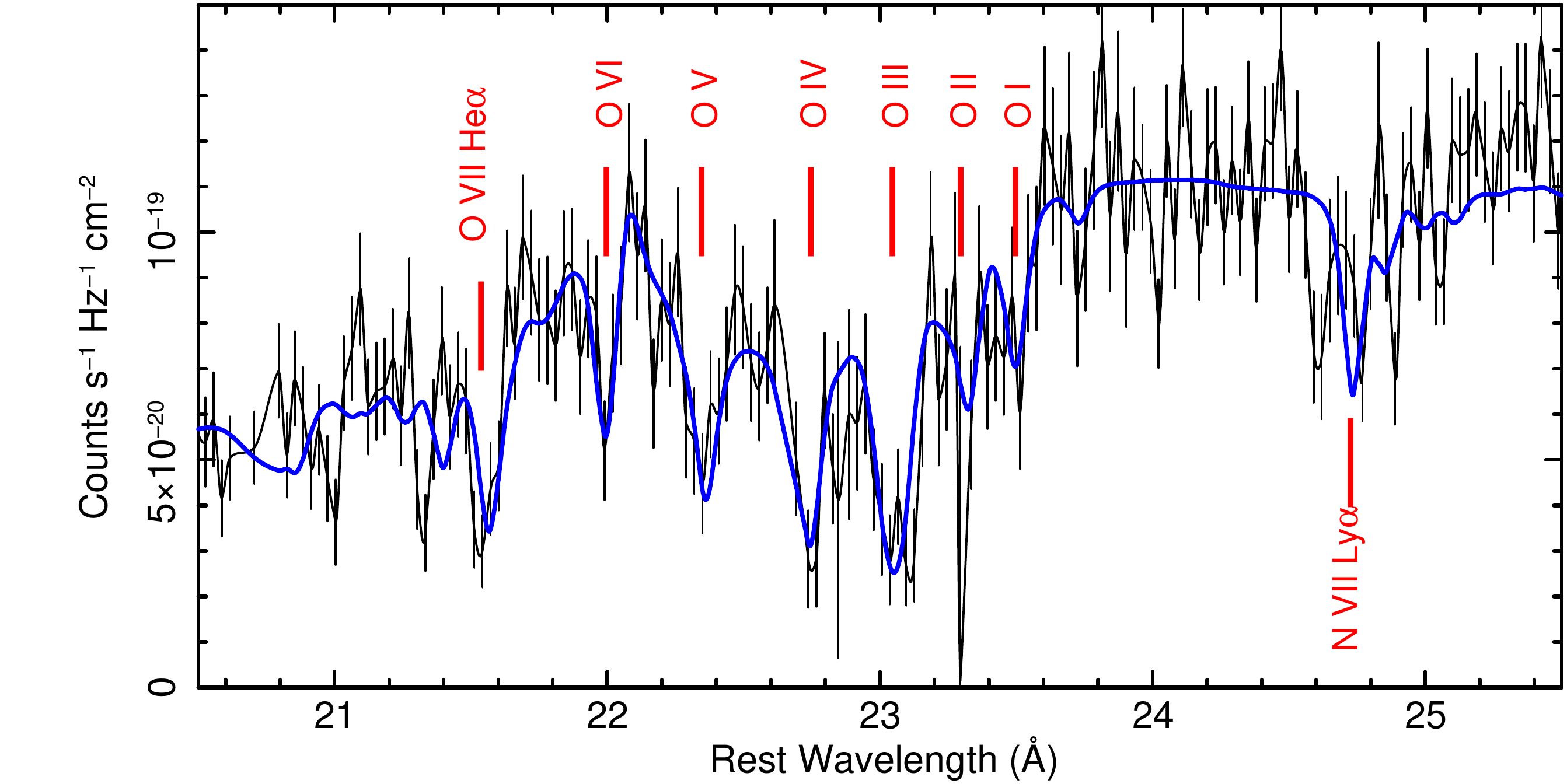}
\caption{A detailed fit to the RGS data from Dec 05.  Black denotes the summed RGS data and the blue line represents the total model.} 
\label{fig:601_UTA}
\end{figure}

\begin{figure}
\includegraphics[scale=0.43,width=6cm, height=8cm,angle=-90]{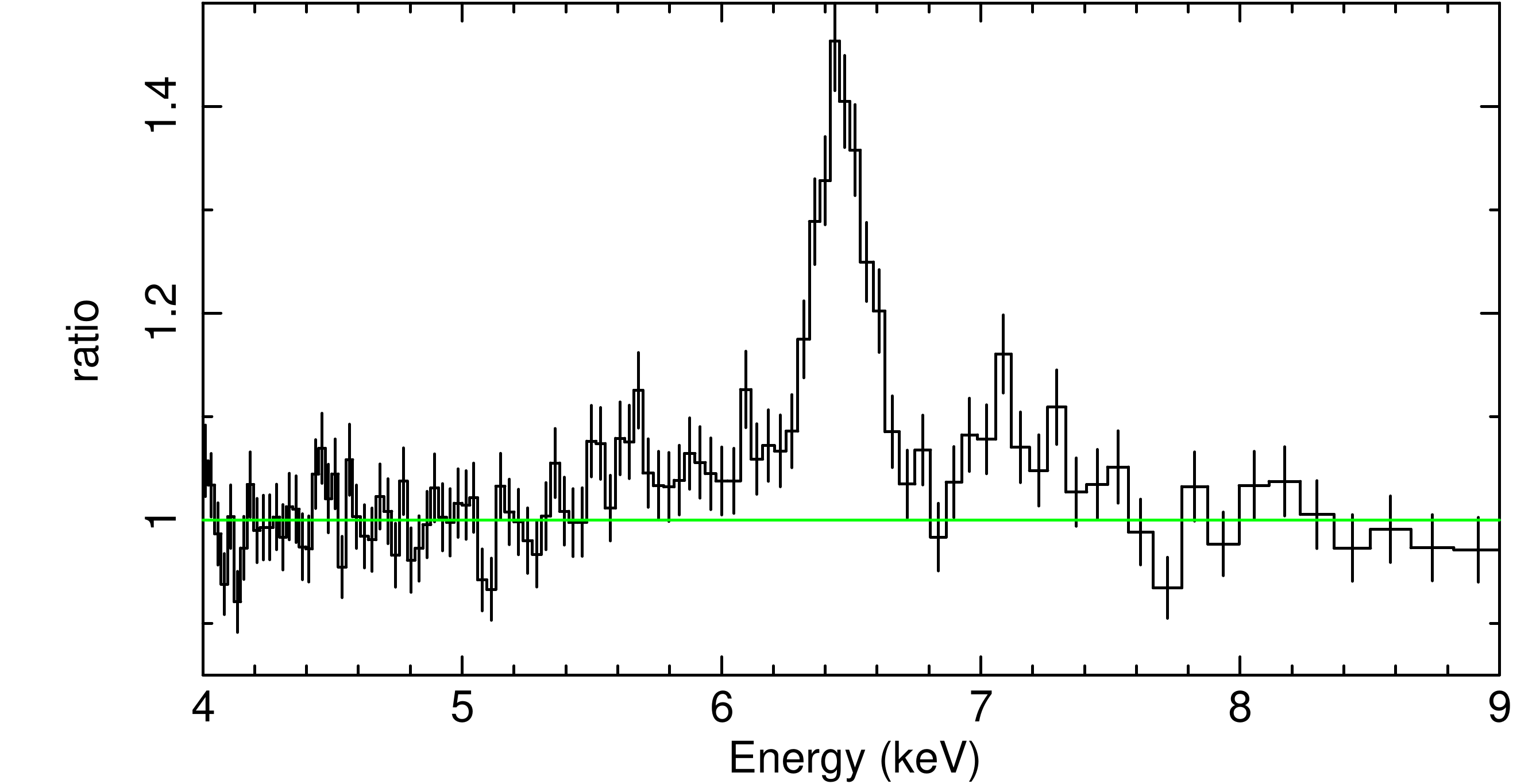}
\caption{The data ratio in the Fe K band, compared to the local continuum model in  pn data for the Dec 05 data.} 
\label{fig:fek}
\end{figure}

\begin{figure}
\includegraphics[scale=0.43,width=8cm, height=8cm,angle=-90]{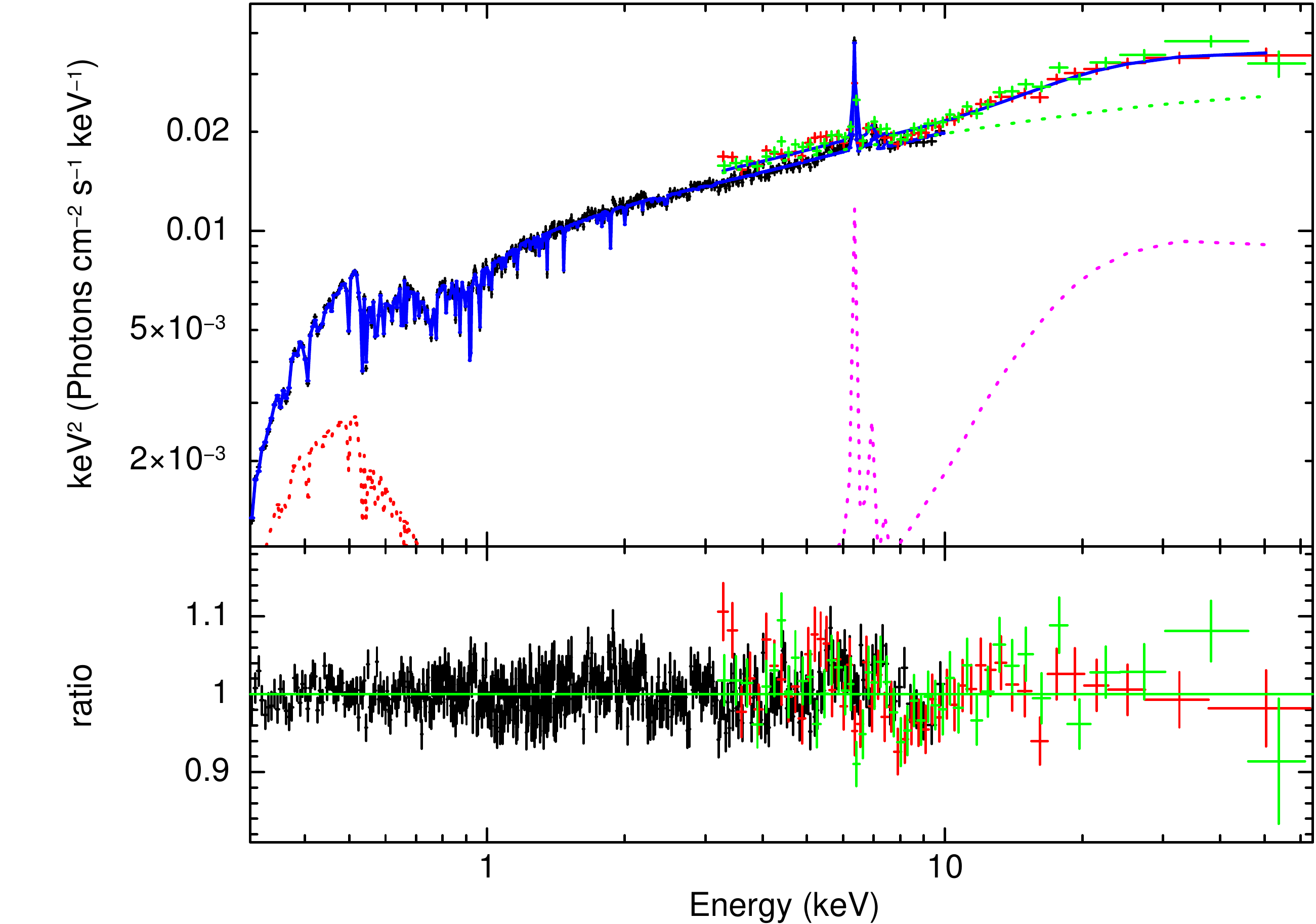}
\caption{Top: a broad band fit (top panel) to the {\it XMM} pn data (black) from Dec 05 plus the simultaneous {\it NuSTAR} FPM data (red and green) compared to the total model (solid blue line). Bottom: the data ratio against the model. } 
\label{fig:601}
\end{figure}

\begin{figure}
\includegraphics[scale=0.43,width=8cm, height=8cm,angle=-90]{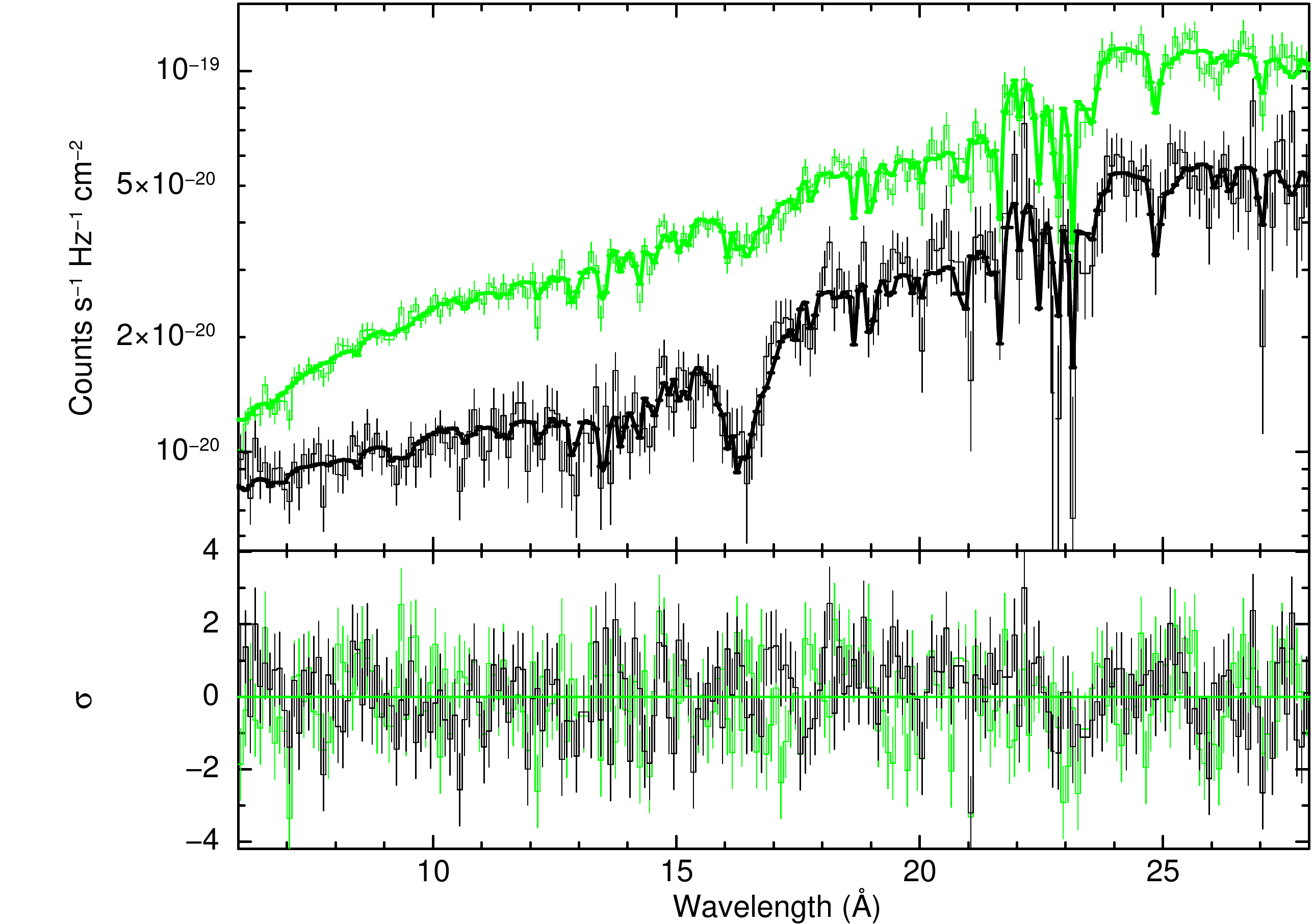}
\caption{A comparison of the warm absorber during observation of Dec 05  (green) versus  the lowest state slice of Dec 09 (black). The solid line shows the model documented in Section 5. The variable UTA feature is evident between 16-17 \AA\ . } 
\label{fig:RGScomp}
\end{figure}

\begin{figure}
\includegraphics[scale=0.43,width=8cm, height=8cm,angle=0]{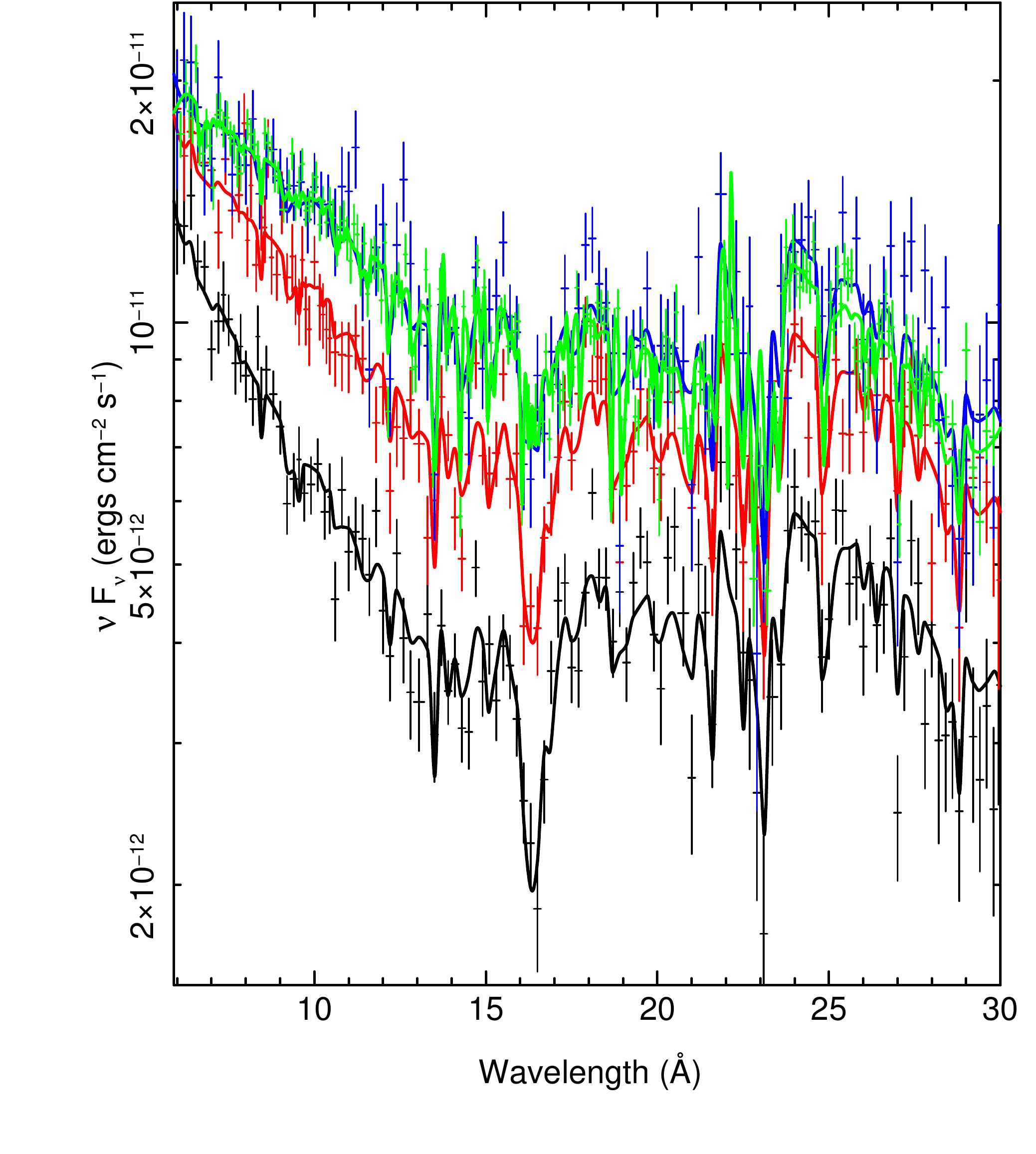}
\caption{Variability in the warm absorber traced by slicing RGS data from the Dec 09 data (slice A is black, slice B is red and slice C is blue) and comparing it to the Dec 05 data (green)} 
\label{fig:RGS_slices}
\end{figure}

\begin{figure}
\includegraphics[scale=0.43,width=8cm, height=8cm,angle=-90]{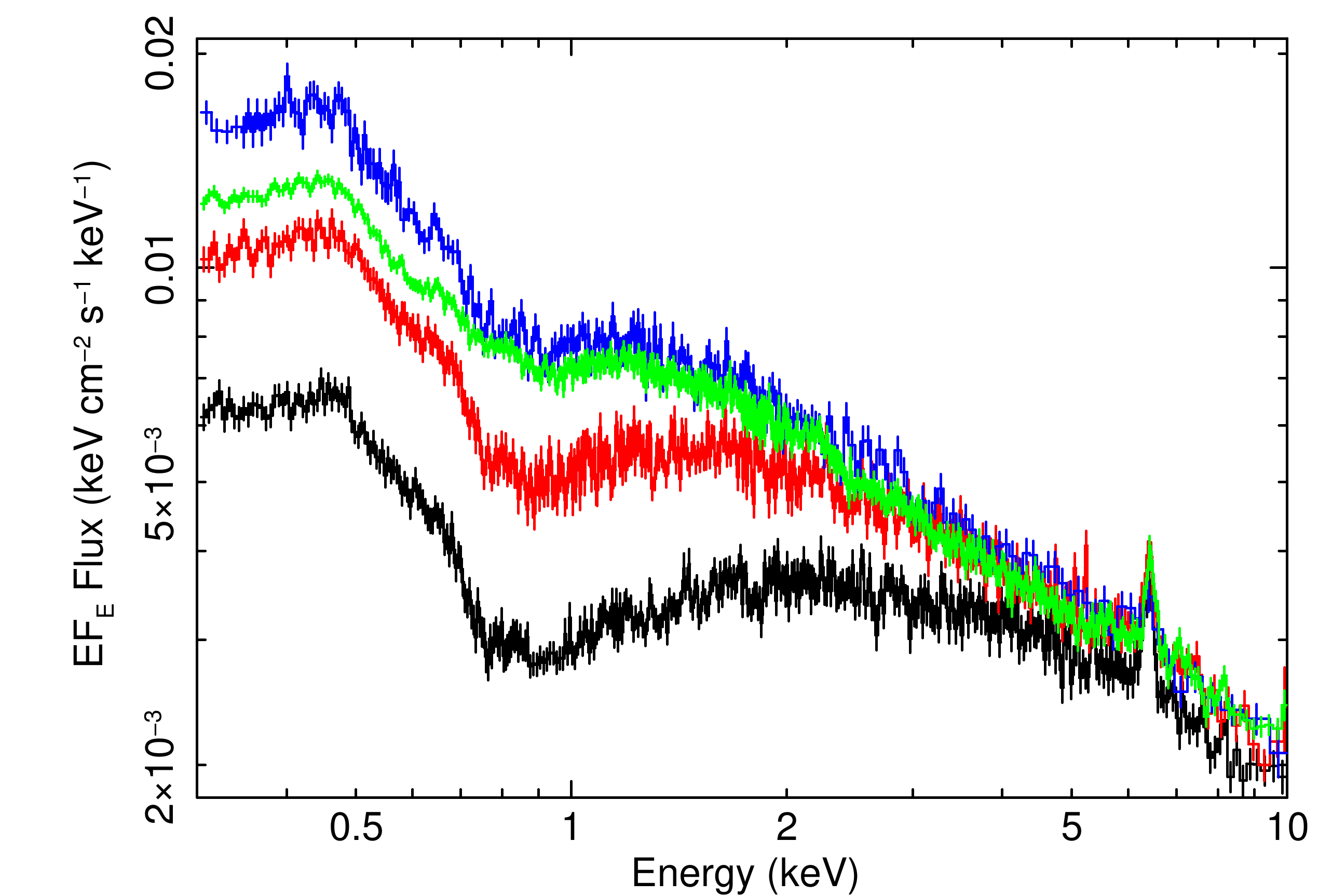}
\caption{{\it XMM} pn spectra: the mean spectrum from the Dec 05 data (green) compared to the slices from  Dec 09, A (black), B (red) and C (blue). The spectra have been unfolded relative to a power-law of photon index $\Gamma=2$. } 
\label{fig:pn_slices}
\end{figure}

\begin{figure}
\includegraphics[scale=0.43,width=6cm, height=8cm,angle=-90]{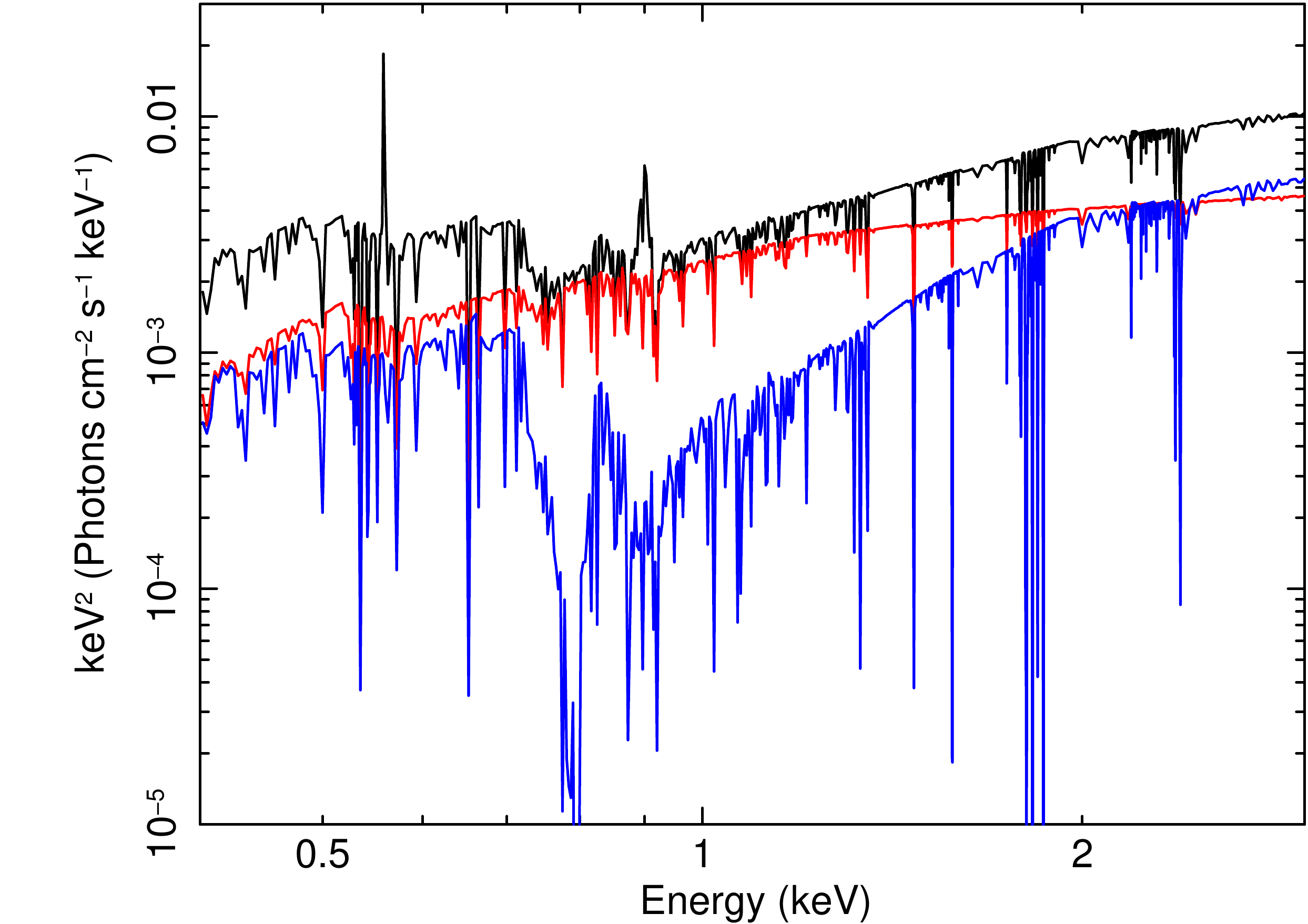}
\caption{The model for slice A of Dec 09 data. 
The blue line is the one absorbed by the heavy pc zone, while the red one is only absorbed by the constant WA (black is the total  model including the emission lines).
The pc zone is essentially adding in the extra opacity around 0.7-0.8 keV and is thus able to model the increase in strength of the UTA feature as well as the curvature of the continuum.} 
\label{fig:3comp}
\end{figure}

\begin{table*}
	\centering
	\caption{RGS fit to 2016 Dec 05: Absorption Complex}
	\label{tab:table}
	\begin{tabular}{|lccc|} 
\hline

Zone & $N_{\rm H}$  & Log\,$\xi$  & Velocity \\  
 &    $\times 10^{21}{\rm cm^{-2}}$  &  & km/s \\
\hline
1& $2.07^{+0.17}_{-0.16}$   & -0.63$^{+0.06}_{-0.57}$   &  $ -157^{+44}_{-46}$   \\
2 (UTA) & 0.83$^{+0.27}_{-0.19}$   & 1.39$^{+0.09}_{-0.11}$ & $ -798^{+47}_{-61}$ \\
 3 & 4.42$^{+1.14}_{-0.95}$ & $2.85^{+0.13}_{-0.09}$ & $- 792 ^{+52}_{-62}$ \\
\hline 
\end{tabular}
\end{table*}

\begin{table*}
	\centering
	\caption{RGS fit to 2016 Dec 05: Line Emission}
	\label{tab:table}
	\begin{tabular}{|l c c c l l |} 
\hline

\hline
 Energy  & $\sigma$  & $\sigma$ & EW  & wavelength & ID  \\
 eV & eV & km/s & eV & \AA\  & \\
 \hline
 561.4(f)  & 0.5(f) & - & $1\pm 1$  & 22.08 \AA\ & O {\sc vii} $f$ \\
564.8$^{+2.9}_{-2.4}$ & $6.9^{+2.4}_{-1.6}$ & 3700 & $7\pm1 $& 21.95 \AA\ & O {\sc vii} $r$ \\
 674.7 $^{+6.9}_{-6.8}$& 21.1$^{+6.2}_{-4.4}$  & 9300 & 8$\pm2$  & 18.38 \AA\  &  O {\sc viii} Ly$\alpha$ \\
911.0$^{+2.4}_{-4.6}$ & 1.9$^{+7.6}_{-1.9}$  & 660 & 4$\pm 1$  & 13.61 \AA\  &  Ne {\sc ix} $f$\    \\
\hline
\end{tabular}
\end{table*}

\begin{table*}
	\centering
	\caption{Spectral variability in the RGS band}
	\label{tab:table}
	\begin{tabular}{ |l l  |}	
\hline

		Seq  &    WA Zone 2    \\
		&   $N_{H}$  $\times 10^{21}{\rm cm^{-2}}$      \\
		\hline
		601      & $1.45^{+0.21}_{-0.28}$   \\
		701a  &  $4.43\pm0.18$   \\
		701b  & $3.03\pm0.27$    \\
		701c    & $1.72\pm0.29$     \\
 \hline
	 
 	\end{tabular}
\end{table*}

\begin{table*}
	\centering
	\caption{Final fit to the X-ray absorption complex after inclusion of the {\bf partial covering (PC)} Zone}
	\label{tab:table}
	\begin{tabular}{|lccc|} 
\hline

Zone & $N_{\rm H}$  & Log\,$\xi$  & Velocity \\  
 &    $\times 10^{21}{\rm cm^{-2}}$  &  & km/s \\
\hline
1 (full)& $1.79^{+0.16}_{-0.19}$   & -0.65$^{+0.06}_{-0.07}$   &   -157(f)  \\
2 (full - UTA) & 1.47$\pm0.23$   & 1.28$^{+0.08}_{-0.13}$ & -798 (f)  \\
 3 (full) & 6.56$\pm 0.27$ & $2.80^{+0.11}_{-0.10}$ & - 792 (f) \\
 4 (PC) & 50.0$\pm2.30$   & 2.23$^{+0.08}_{-0.13}$ & -798 (f)  \\
 & & & \\
\hline
\end{tabular}
\end{table*}

\begin{table*}
	\centering
	\caption{Covering changes in the PC zone for the final fit}
	\label{tab:table}
	\begin{tabular}{ |l l |M{1.5cm} |}	
\hline
		 Seq  & PC Zone Covering  Fraction  \\
		\hline
		601       & $0.00 (<0.11)$  \\
		701a   & 0.64$\pm0.09$ \\
		701b  & $0.39\pm0.02$  \\
		701c       & $0.00(<0.21)$   \\
 \hline
	 
 	\end{tabular}
\end{table*}

\section{Initial Overview of the Campaign}

Light curves were constructed and the individual sequences from the campaign concatenated together (to remove the gaps, for visual clarity regarding the source behaviour).  NGC\,3227 shows strong X-ray variability across the observation set
(Figure ~\ref{fig:pn_lc})  across the bandpass of the {\it XMM} data. Initial inspection of the data revealed the source spectrum to be hard  in
the low flux states, and soft in the high states, as observed
previously for local Seyfert galaxies \citep[e.g.][]{miller10a}. Consideration of the hardness ratio behaviour allowed the 
isolation of an epoch of unusual spectral behaviour, i.e. an abrupt hardening of the source spectrum at the start of OBSID 0782520701, the exposure from 2016 Dec 09, which was the last observation of the 2016 campaign. 
 Figure~\ref{fig:hardness} shows the significant hardening and illustrates how that is different in characteristics to  previous sequences in  this campaign, even those where NGC 3227 occupied a similar flux state. 
 
To further our understanding we constructed a light curve from the {\it NuSTAR} data  over the 10-50 keV band, and compared this with the {\it XMM} pn and RGS light curves over the Dec 09 data (note that there was a slight offset between the start times of the {\it NuSTAR} and {\it XMM} segments of  data).  This comparison showed the 10-50 keV band to be steady in flux, while the 
lower energy flux levels varied (Figure ~\ref{fig:pn_rg}).  Based upon the variability of flux below 10 keV, we split the data from Dec 09 into three intervals to sample the source spectrum as the flux changed.  The time intervals chosen are shown in Figure~\ref{fig:pn_rg} and denoted by the vertical dash-dot lines. 
We sub-divided the spectral data and examined the shape changes across the three time slices chosen: Slice A was the first segment of the Dec 09 data and had an exposure of 27 ks, slice B was the middle segment with an exposure 17 ks, and slice 3 the final segment with just 6 ks (although being the highest flux state this had adequate counts for analysis).  These slices will be considered in the spectral variability analysis presented in Section 5. 

\section{Characterization of the baseline X-ray spectrum}

Motivated by the sudden change in spectral behaviour exhibited on Dec 09, we proceeded to parameterize the  X-ray form during 2016 Dec 05, to establish the spectral baseline immediately prior to the hardening event.

\subsection{Characterizing  Dec 05 RGS data}

We began with a detailed analysis of the RGS spectral data from Dec 05, to fully characterise the warm absorber complex. 
Owing to the narrow bandpass of the RGS and the complexity of the absorbing gas transmission profile, it is not possible to also determine the underlying spectral continuum from those data. We therefore assumed a photon index of 1.8 and allowed in the fit a black body component (that improved the fit from  $\chi^2=1316/835\, d.o.f.$ to $\chi^2=1033/833\, d.o.f.$ in the final best fit) whose temperature was 
$kT=90$ eV: these values were obtained from a fit to the {\it XMM} pn and {\it NuSTAR} data and together they model the underlying continuum. We return to the pn fit in the next section.

The RGS spectrum shows numerous absorption features, and some emission lines, covering a wide range of ionization states of material. It was necessary to include three {\sc xstar} zones of ionized gas in the model, to adequately model these data.  The gas zones showed no evidence for partial covering in the RGS data and so were all assumed to fully cover the source.  

The spectral data from the combined RGS gratings are shown in Figure~\ref{fig:601_UTA}. All of the warm absorber zones have column densities in the range $10^{21} - 10^{22} {\rm atoms\, cm^{-2}}$ and cover a wide range of ionization (Table~2). 
Zone 1 is of relatively low ionization (Table~2) and is responsible for the slew of inner K-shell absorption lines from O {\sc i} - {\sc vi} between 20 - 25 \AA\, whose wavelengths indicate a low bulk outflow velocity $\sim 150$ km/s for this gas.  Zone 2 most notably shows a broad signature of Fe M-shell absorption, in an unresolved transition array (UTA) centered between 16 - 17 \AA\, from gas outflowing at  $\sim 800$ km/s. 
Zone 3 has strong absorption lines from highly ionized species, such as the H-like lines from N {\sc vii} Ly$\alpha$, O{\sc viii} Lyman $\alpha$ and Lyman $\beta$. This zone shows the same outflow velocity as Zone 2. 

The fit left significant positive residuals around $\sim$ 18.4 \AA\,(0.674 keV), where an O {\sc viii} Ly$\alpha$ emission line is expected, and so we added a broad Gaussian line component to the fit to account for this. Excess emission is also associated with the O 
{\sc vii} triplet near 22\,\AA, where we included a weak narrow component corresponding to the forbidden line and a broader emission component that could correspond to a blend of line emission from the intercombination and resonance line components. He-like emission is also present at 13.6\,\AA\, (0.91 keV), likely  from Ne {\sc ix}.  This final fit gave $\chi^2=1033/833\, d.o.f$, compared to $1117/843$ without the emission lines. Table~3 shows fitted line energy, widths, velocity width, equivalent width against the local continuum, fitted wavelength and line identification.  
A future paper will discuss a detailed fit  to the full set of  RGS data from this campaign, including the emission line profiles from the mean spectrum.

\subsection{Fitting the broad X-ray spectral form for Dec 05}

Having established a model for the signatures of X-ray absorbing gas in the 0.4--2.0 keV band,  we then performed a joint fit to the {\it XMM} pn  data from Dec 05 and simultaneous {\it NuSTAR} FPM data from 2016 Dec 05.  A model was constructed composed of a power-law continuum modified by passage through  ionized gas. As the pn has a lower spectral resolution than the RGS, and the data were taken simultaneously, we fixed the parameters of the three absorbing zones of gas at the values already determined from the RGS fit (Table~2). 
Reflection was added into the model, based on the presence of a weak ($\sim 70$ eV) and narrow  Fe K$\alpha$ emission line at 6.4 keV (Figure~\ref{fig:fek}). A reflection signature would be expected from gas out of the line-of-sight, such as might arise from a geometrically thin, optically thick disc subtending $2\pi$ steradians to the continuum source. Reflection was parametrized for neutral gas, using the {\sc xspec} model {\sc pexmon} and the illuminating photon index was linked to that of the continuum, this component provided a good parameterization of the  Fe K$\alpha$ emission in the spectrum, for an Fe abundance of $1^{+0.3}_{-0.2}$ and reflection fraction $R =1.07\pm0.16$

The best-fit model  to the mean 2015 Dec 05 spectrum  
yielded a photon index $\Gamma=1.84\pm0.01$ with a high-energy cut-off  $E_{cut}=309\pm73$ keV.  
The soft part of the spectrum  was parameterized by a black body component with $kT=90 \pm 3$ eV, as mentioned previously. 
This model provided a good fit to the broad X-ray form (Figure~~\ref{fig:601}), with $\chi^2=2215/2035\, d.o.f$.

To estimate the ionizing luminosity,  we added the {\it XMM} OM UVW1 filter point to the fit. We accounted for a reddening of E(B-V)=0.18 in the model, using the {\sc xspec} model {\sc redden}, based upon the results of \citet{crenshaw01a}.  Extending the $\Gamma=2$ powerlaw down to the UVW1 band, we estimate the integrated luminosity over 1 - 1000 Ryd, after correcting for all intrinsic absorption,  to be $L_{ion} \sim 8 \times  10^{42} {\rm erg\, cm^{-2}s^{-1}}$.

\section{Spectral Variability}

As the RGS data provide a sensitive test of absorption changes, we first compared the summed RGS data from Dec 05 (601) with the lowest state taken during Dec 09, i.e. 701 slice A.
Figure~\ref{fig:RGScomp} provides an immediate visual confirmation of a significant change in the depth of the UTA feature during the low state comprising slice A, compared to the prior 601 sequence. This requires at least a factor of $\sim 3$ increase in the column density of Zone 2 (Table~4), to give an occulting column density of $N_{\rm H} \sim 4.4\times 10^{21} {\rm cm^{-2}}$ during slice A of the UTA zone of material having an ionization parameter of $\log\xi \sim 1.4$ (Table~2). 

We then extended the analysis to compare all three RGS slices from Dec 09 (701) with the mean (601) spectrum from Dec 05. 
Allowing only the column density of Zone 2 to be variable provides sufficient change in the RGS band to account for the spectral variability observed (Figure~\ref{fig:RGS_slices}). The best-fitting parameters for each zone are shown in Table~4, where the column density of the UTA zone 
increases from $\sim N_{\rm H}=1.5 \times10^{21}$\,cm$^{-2}$ in the 601 spectrum to a maximum of $N_{\rm H}=4.4 \times10^{21}$\,cm$^{-2}$ 
in 701 slice A, which then declines by a factor of 3 within a timescale of 100\,ks by the end of slice C. Thus the observations appear to capture a rapid increase and then decline in the soft X-ray absorber in NGC 3227.
Note that in comparison, there is no requirement for any significant variability of either the low ionization zone 1 or the high ionization 
zone 3; all of the absorber variations in the RGS band appear to be associated to the UTA zone 2.

Next, in order to build up a broad band picture of the spectral variability, 
we compared the Dec 05 pn spectral data, with the time-sliced Dec 09 pn spectra. Inspection of those data shows marked variation in curvature in the few keV regime across the time periods of interest (Figure~\ref{fig:pn_slices}).  In contrast to this, as previously noted, the source is steady above 10 keV  (Figure~\ref{fig:pn_rg}), i.e the spectra converge to high energies and no significant spectral variations are seen in any of the 
{\it NuSTAR} spectra above 10\,keV across either sequence 601 or 701.
This allowed us to proceed using a photon index and reflection contribution that are linked (but not frozen) to the same values for  the Dec 05 spectrum and the Dec 09 spectral slices, although small variations in the normalizations of the continuum (powerlaw plus blackbody) 
components were allowed between the slices to account for any intrinsic variability.\footnote {Small changes in the black body flux were measured at the level of $\sim \pm 5\%$ between the slices. As these cannot be degenerate with the absorber changes, that are based upon strong absorption lines in the RGS band, these fluctuations are not tabulated.}

After applying the successful RGS model to the time-sliced pn data, we found that a change in column density for the full covering zone 2 is no longer sufficient -- as, with the pn data included, we need to explain both the change in depth of the UTA and the strong spectral curvature variations, which is especially noticeable between the Dec 05 spectrum and slice A  from Dec 09 (Figure~\ref{fig:pn_slices}).  Indeed the fit allowing only the 
zone 2 fully covering absorber to vary is very poor, with $\chi_{\nu}^2=7567/4818$, as the variability of this UTA zone alone is not 
sufficient on its own to reproduce the strong spectral curvature seen at the start of the 701 pn sequence.

Therefore we added another absorber zone to the model (zone 4), of higher column density, using an {\sc xstar} table configured to partially cover the continuum.  Thus, the model form was:

$$N_{H,Gal} \times Z_1 \times  Z_2  \times Z_3 [( 1 - f) \times (po + ref  + bb) + f \times Z_4 \times (po + bb)]$$

where Z 1,2,3 and 4 correspond to the ionized absorber zones of Table~5. $N_{H,Gal}$ is the Galactic column represented by {\sc tbabs}, po is the power law, bb, the black body, ref the reflector represented by {\sc pexmon} and f is the covering fraction for zone 4.

The outflow velocity for the partial covering (or ``PC'') zone 4 (that is not determinable from the lower resolution pn data) was assumed to be equal to that of zone 2 determined from the RGS spectra (and which is also equal to that of Zone 3). The fit yielded provided a good model to all the pn spectra, with $\chi^2= 5346/ 4811 \, d.o.f$.  The PC Zone was found to have $N_{H(PC)}=5.00^{+0.31}_{-0.06} \times 10^{22} {\rm cm^{-2}}$ and log $\xi=2.22^{+0.04}_{-0.04}$. 
Here we allowed variations in the covering fraction of the PC zone to account for the opacity changes across sequences 601 and 701.  
The PC Zone {\bf is also a UTA producing zone}, i.e. it gives significant opacity in the 0.7--0.8 keV regime at the UTA, as can be seen in 
Figure~\ref{fig:3comp} (blue curve), as well as producing significant spectral curvature in the soft X-ray band.  

Thus the change in covering of the PC zone can simultaneously account for the increase in the soft X-ray spectral curvature of slice A  and the increase in strength of the UTA feature, {\bf without recourse to additional changes in the column of the full covering zone 2}.   The column densities and ionization parameters of the full covering zones were therefore linked between slices for the final fit and the outflow velocities were frozen at the values derived from the RGS analysis (see Table~5 for final values). The covering changes of zone 4, that are all we require to explain the absorption event, are shown in Table 6. 

Figure~\ref{fig:3comp} shows the final absorber model components for slice A of  the Dec 09 data, where the 
 blue line is the fraction of continuum absorbed by the large-column partial-covering zone, while the red line shows transmission through the  constant 
 part of the warm absorber complex (black shows the total spectrum including the emission lines).  The UTA is evident in the partial-covering zone, and we note that there is absorption from a blend of O {\sc vii - viii} edges present adjacent to that.

\section{Discussion}

We have analyzed new data from simultaneous {\it XMM} and  {\it NuSTAR} observations of NGC 3227 during 2016 and found 
a complex X-ray absorber that we have modeled as four zones of ionized gas. Three zones of the absorber are consistent with fully covering the source, while changes in covering fraction of the forth, PC Zone, account for marked rapid spectral variability.
 
Following \citet{reeves18a} we consider the radial locations of the zones, based upon the fitted spectra and variability timescales, using the simple relationships that govern the gas behaviour:
 
\begin{enumerate}

\item  $\Delta r = v \times \Delta t$ where $v$ is the velocity of the cloud, existing with size $\Delta r$ and causing an occultation event of 
duration $\Delta t$ 

\item $v^2=\frac{GM}{r}$ that assume the cloud to be in a Keplerian orbit, at a radial distance $r$ from the black hole of mass $M$

\item $n_e= \frac{\Delta N_H}{\Delta r}$ where $n_e$ is the electron density (equivalent to the particle density), $\Delta N_H$ is the change in column density observed, ie the integrated column density of the occulting object

\item $r^2=\frac{L_{ion}}{n_e \xi}$ (see Section 2.3 for quantity definitions)

\vspace{4mm}
Simultaneous consideration of  i - iv allows algebraic substitutions to be made that negate the need to assume or estimate $\Delta r$ or $n_e$, giving
\vspace{2mm}
\item  $r^{\frac{5}{2}}=(GM)^{\frac{1}{2}} \frac{L \Delta t}{\Delta N_H \xi}$

\end{enumerate}

NGC 3227 has a black hole of mass $M_{BH}= 5.96^{+1.23}_{-1.36} \times 10^6 M_{\odot}$  \citep{bentz15a}, with corresponding gravitational radius  $r_g \sim 10^{12}$ cm. 
Considering the change in column density affected by passage of clumps of gas from the PC Zone, we take the timescale for the occultation event to be $\Delta t \sim 10^5 s$, for the PC Zone the clouds provide  $\Delta N_H \sim 5 \times 10^{22} {\rm cm^{-2}}$ and $\xi \sim 166$. 
 In Section 4.2 we estimated $L_{ion}$ to be  $\sim 8 \times 10^{42} {\rm erg\, s^{-1}}$.

Substituting these values into (v) yields $r \sim 6 \times 10^{15}$ cm.  Taking this radial estimate, we can estimate the gas transverse velocity from (ii) to be  $v \sim 4000$ km/s and then the cloud size from (i) to be $\Delta r \sim 4 \times 10^{13}$ cm. Following \citet{elitzur06a} and \citet[][their eqn. 14]{beuchert15a} we can  check the viability of this cloud size by examining the upper limit on the size of a cloud that can withstand tidal shearing, finding $\Delta r < 2 \times 10^{13}$cm. Thus, our cloud size estimate is reasonable. Finally, we can use these derived values in (iii) to obtain  an estimate of the gas density for Zone 2, $n_e \sim 10^9 {\rm cm^{-3}}$. 

The optical broad line region in this AGN is estimated to exist at a radius of 10-20 light days \citep{salamanca94a}, i.e.  $\sim 2 - 5 \times 10^{16}$ cm. The radial estimate for the  PC Zone lies within the BLR, so this gas may comprise or be associated with the clouds of the inner BLR.   The timescale for this variation, of the order of a day, is similar to X-ray absorption variations discovered in some other AGN whose clouds have also been suggested to lie close to the BLR, e.g. NGC 3516 \citep{turner08a},  NGC 1365 \citep{braito14a},  NGC 3783 \citep{mehdipour17a} and PG1211$+143$ \citep{reeves18a}. The comparison between NGC 3227 and PG1211$+143$ is especially compelling as in that case the spectral changes are also associated with variations in the UTA feature.

It is interesting that  Zone 3, the PC Zone and Zone 2 all have consistent outflow velocities.   A plausible picture is that Zone 2, 3 and the PC zone are all part of the same complex cloud structure. Zone 3 may represent the absorption from the limbs of the PC clouds. Clouds, by their geometry, offer a smaller column density at their limbs than through the center. The smaller limb-column would naturally suffer a greater degree of ionization by the continuum, consistent with the measurements.

 Finally, Zone 1 has a column density $\sim 2 \times 10^{21}\, {\rm cm^{-2}}$ of cool gas (Table~5) outflowing with a velocity $\sim 150$ km/s.   The UV continuum of  NGC 3227 is heavily reddened \citep{komossa97a} with a column density for the dusty, reddening gas  estimated to be  
$> 2 \times 10^{21}\, {\rm cm^{-2}}$ by \citet{kraemer00a}, an estimate that was supported by  \citet{crenshaw01a}.   The  NLR gas shows a maximum velocity  of $\sim 500$ km/s \citep{fischer13a} in this source.  
 The radial location of the NLR is  $\sim 10^{20}{\rm cm}$ \citep{schmitt96a} in NGC 3227,  existing outside of the 
dust sublimation radius  (estimated to be $\sim 10^{17}$cm, \citealt{beuchert15a}).  The properties of Zone 1 are consistent with arising as part of the NLR gas and in  such a picture, this zone would therefore lie outside of the other components of X-ray absorption in NGC~3227.


\section{Conclusions}

We have performed detailed time-resolved spectroscopy of a rapid spectral variability event, occurring over approximately one day, toward the end of a month-long monitoring campaign using {\it XMM-Newton} with {\it NuSTAR}. RGS  data reveal a 
 a UTA imprinted on the spectrum by  the X-ray absorber complex.  An increase in the UTA depth during the source low state on 2016 Dec 09  identifies the rapid spectral variability event as changes in absorption.  The data are consistent with transit of a gas cloud  having $N_H \sim 5 \times 10^{22} {\rm atoms\, cm^{-2}}$, log $\xi \sim 2$, whose movement into the line-of-sight occults 60\% of the X-ray continuum photons during the start of the Dec 09 exposure.  
 The occulting cloud  is estimated to exist on the inner edge of the optical BLR, and has an outflow velocity $\sim 800$ km/s. One of the additional warm absorbing layers  matches the outflow velocity of the variable zone, and may represent transmission through the cloud limb. 

 NGC 3227 is a complex and heavily absorbed source, whose absorption feature variability make it  a high priority for examination with the forthcoming micro-calorimeter on The X-ray Imaging and Spectroscopy Mission ({\sc XRISM}).

\section*{Acknowledgements}

TJT acknowledges NASA grant NNX17AD91G.  Valentina Braito acknowledges financial support through NASA grant NNX17AC40G and through the CSST Visiting Scientist Initiative.  APL acknowledges support from STFC consolidated grant ST/M001040/1. We are  grateful to the {\it XMM} and {\it NuSTAR} operations teams  for performing  this campaign and providing 
software and calibration for the data analysis.  




\bibliographystyle{mnras}
\bibliography{xray_2018} 





\bsp	
\label{lastpage}
\end{document}